# The GenAI Who Knew Too Little – Revisiting Transactive Memory Systems in Human GenAI Collaboration

**Research Paper**

Christian Meske[1], Tobias Hermanns[1], Florian Brachten[1]

[1] Ruhr-University Bochum, Chair of Socio-technical System Design & AI, Bochum, Germany
{christian.meske,tobias.hermanns,florian.brachten}@rub.de

**Abstract.** Generative artificial intelligence (GenAI) is rapidly transforming knowledge work, creating new collaborative dynamics that challenge traditional human-human interaction patterns. This study investigates how Transactive Memory Systems (TMS) manifest in human-GenAI collaboration. Through qualitative interpretive interviews with 14 knowledge workers and thematic analysis, we identified distinct TMS manifestations across its three main dimensions: (1) regarding specialization, expertise recognition exhibited asymmetry. Humans developed awareness of GenAI capabilities while GenAI demonstrated only temporary recognition of human expertise; (2) in terms of credibility, humans formed ambivalent trust relationships with GenAI, simultaneously attributing expertise to the latter while implementing additional verification processes; and (3) concerning coordination, where humans primarily directed more passive GenAI systems lacking initiative, resulting in hierarchical interaction patterns and unidirectional information flow. Our findings reveal that TMS manifests differently in human-GenAI compared to human-human collaboration, necessitating theoretical adaptations including trust markers, acknowledging GenAI's absent metamemory, and identifying its temporary memory.

## 1 Introduction

Generative artificial intelligence (GenAI) has experienced substantial growth in recent years, reshaping the landscape of knowledge work and introducing the potential for new collaborative dynamics that extend beyond traditional human-to-human interactions (Torkamaan et al., 2024; Söllner et al., 2025). With increasingly powerful language models, GenAI has been rapidly integrated into numerous professional workflows, evolving from simple tools to sophisticated collaborative agents (Lu et al., 2024). Thus, GenAI can serve as partners in complex processes, offering assistance across a various set of knowledge tasks such as text generation, programming, information retrieval, or creative ideation (Ozkaya, 2023; Huynh, 2024; Ruta et al., 2025).

These novel arrangements pose new challenges from a theoretical perspective, as knowledge workers paired with GenAI exhibit different interaction and information processing patterns (Ozkaya, 2023) than human collaborators that rely on social dynamics and curated knowledge about each other in order to succeed in collaborative knowledge tasks (Lewis 2003; Lewis & Herndon, 2011).

Transactive Memory Systems (TMS) theory captures these underlying socio-cognitive processes and structures within collaboration settings. Originally conceptualized by Wegner et al. (1985, p. 256) as “a combination of individual minds and the communication among them”, TMS explains how humans collaborate based on a shared awareness of ‘who knows what’ within teams. This structure manifests through three behavioral dimensions: *specialization* (recognition of other’s expertise), *credibility* (trust in others' expertise), and *coordination* (effective task division) (Lewis & Herndon, 2011). TMS theory has proven helpful for explaining human-human collaboration (e.g., Alahmari et al., 2023; Hopf et al., 2024) and has been leveraged in Information Systems research in various contexts, such as virtual teams (Chen et al., 2020), distributed IT projects (He et al., 2022), and management (Donate et al., 2023).

As GenAI increasingly integrates into knowledge work, it challenges TMS theory which was developed  and utilized to explain collaboration solely among humans (Qu et al., 2023; Restrepo-Tamayo et al., 2024). A critical tension emerges because while TMS theory relies on socio-cognitive processes built through human social interaction, GenAI functions through algorithmic patterns without these social mechanisms. This creates a theoretical gap where purely social frameworks cannot adequately describe what has become a socio-technical relationship. Therefore, empirical examination of human-GenAI collaborations in practice is essential to determine which aspects of TMS persist, transform, or disappear in these novel work arrangements.

Hence, our study addresses the following research question: *How do TMS manifest in human-GenAI collaboration, and how can a nuanced perspective on TMS represent these new socio-technical relationships?*

To investigate our research question, we conducted qualitative interpretive interviews with 14 knowledge workers who regularly engage GenAI for knowledge tasks. A thematic analysis (Braun & Clarke, 2006) revealed distinct TMS manifestations in human-GenAI collaboration across the three key TMS dimensions: (1) *specialization*, where humans recognized GenAI’s capabilities, but GenAI only temporarily recognized human expertise; (2) *credibility*, where humans exhibited ambivalent trust, simultaneously attributing expertise to GenAI while also seeing the need to verify its results; and (3) *coordination*, where humans led interactions with passive GenAI, resulting in unidirectional information flow and engagement. These findings suggest traditional TMS structures must be adapted, as humans develop new trust markers in metamemory, while GenAI operates with temporary rather than persistent memory that for aid in collaboration.

The remainder of this paper is organized as follows. First, we examine literature on human-AI collaboration and we present the theoretical background on TMS (Section 2). Next, we detail our research methodology (Section 3), followed by a presentation of our findings (Section 4). We then discuss theoretical and practical implications, and propose a nuanced view to TMS theory for human-GenAI collaboration (Section 5). This paper ends with a conclusion, including limitations and an outlook to further research (Section 6).

# 2 Background

## 2.1 Human-(Gen)AI Collaboration

Human-AI collaboration occurs when humans trigger AI systems for support, AI systems prompt human evaluation, or both act complementarily in ‘hybrid intelligence’ (Dellermann et al., 2019). Studies indicate that the acceptance of AI depends on one's domain expertise (Jussupow et al., 2021) and perceived self-efficacy (Jussupow et al., 2022). While AI-human disagreement seems problematic, Abdel-Karim et al. (2023) found that discrepancies can trigger deeper reflective processes, potentially enhancing effectiveness. A study by Fügener et al. (2021) found that AI advice increases human accuracy but simultaneously decreases unique human knowledge, indicating a tendency that humans may lower their effort once AI is integrated into collaborative tasks. The effect could in part be mitigated by adapting AI systems to individual human work patterns and their expertise. Yan et al. (2024) highlight engaging with GenAI can achieve better results but once GenAI support is removed, result quality descreases, suggesting a dependency that can reduce skill development. Their findings reinforce concerns and highlight how GenAI literacy may become a crucial aspect for maintaining cognitive autonomy. The authors thus propose a human-centered approach to GenAI collaboration that focuses on developing complementary skills.

Rowe et al. (2024) further highlights the ‘irony of automation’ where systems designed to relieve users from tasks may actually demand more effort as users must ensure proper system functioning. As trust becomes an important factor, questions remain about when AI should take control, how this should be communicated, and what users expect when delegating tasks to AI. ‘Algorithm aversion’ further describes the tendency to avoid AI advice despite its general superiority (Dietvorst et al., 2018). This aversion can be overcome when humans maintain some control over decisions. Logg et al. (2019) found people generally prefer algorithmic over human advice, though resistance may come from overconfidence in personal judgment. An apparent difference existed between professionals and lay people with the latter being more open to algorithmic suggestions while the former were more likely to rejected recommendations and relied on their knowledge. Celiktutan et al. (2024) further reveal a bias in GenAI collaborations, where individuals evaluate their own contributions to work co-produced with GenAI as higher than they evaluate others'. They explained that bias stems from differential assumptions about the collaborative behavior. People believe they engage GenAI primarily for inspiration while assuming others use it to just outsource work. These perception differences directly influence acceptability judgments, where people consider the integration of GenAI into their workflows more acceptable for themselves than for others, posing challenges for trust within teams.

Regarding impression management, human users interacting with generative AI appear more open and less inclined to manage impressions compared to human-human interactions (Gnewuch et al., 2023; Siemon, 2023). This might increase self-disclosure and meta-knowledge within teams. However, Alavi et al. (2024) caution that organizational deployment of GenAI may lead to overreliance on AI for knowledge management, potentially impeding tacit knowledge and reducing human interactions.

When examining AI systems members in larger teams rather than in dyadic human-AI configurations, Dennis et al. (2023) found no algorithm aversion in teams and no difference in willingness to work with AI versus human team members. AI team members were perceived as more objective and potentially helpful in resolving team conflicts. Trust appears to depend more on team performance than on member type (McNeese et al., 2021). Team size is also relevant, with smaller groups benefiting from AI while larger teams' performance may decline due to reduced unique human knowledge (Fügener et al., 2021). Well-performing AI is more likely to be perceived as an actual team member compared to lower-performing AI (Hopf et al., 2024).

For successful human-AI collaboration, Fügener et al. (2022) identify three necessary conditions: existence of complementarities between human and AI skills, recognition of these complementarities, and execution of efficient delegation rules. TMS theory thus offers a valuable framework for understanding these emergent collaborative patterns by examining how expertise is recognized, trust is managed, and work is coordinated between humans and GenAI.

## 2.2 Transactive Memory Systems

TMS were originally conceptualized as systems that combine individual cognition with interpersonal communication (Wegner et al., 1985) in a knowledge sharing and application context. They describe the socio-cognitive structures and mechanisms (Aissa et al., 2022) by which groups cooperatively learn, store, and access knowledge on a group-level (Wegner, 1987; Hollingshead, 2001). In TMS, each group member doesn't need to recall relevant information alone while accomplishing a task. Instead, individuals in a group have unique knowledge and draw on each other as necessary. Wegner et al. (1985) and Wegner (1987) thus conceptualized TMS as individuals' interconnected memories that collectively hold knowledge, supported by specific cognitive processes for accessing and utilizing that information.

In TMS, personal memory describes an individual's cognitive system that stores and organizes information. It encompasses knowledge (facts, experiences, skills) and organizational labels for categorization of this knowledge and access. Memory function relies on three processes: encoding (initial processing), storage (retention), and retrieval (access) (Wegner et al. 1985; Wegner 1987). Personal memory also includes metamemory structures that track what one knows and doesn't know. Metamemory additionally extends cognition to external storage systems (like books or notes) by recording information labels and locations rather than the information itself, enabling retrieval from these external repositories (Wegner et al. 1985; Wegner 1987).

TMS further extend these cognitive principles to the group level, creating a distributed network (Wegner, 1995). Like individuals using metamemory to track archival knowledge locations, groups utilize their individuals' metamemories to track and utilize expertise across group members. Personal memories function as interconnected components within a larger system rather than in isolation. Each member's unique knowledge becomes a specialized entity in the group's collective coordination architecture. Individual metamemory processes expand to track others' expertise domains, creating what Wegner (1995, p.326) called a directory of "who knows what".

Transactive processes are mechanisms through which group members coordinate knowledge creation, storage, and retrieval from each other for effective task application in cooperative settings. These include encoding (learning expertise distribution), storage (maintaining awareness of who knows what), and retrieval (accessing information through appropriate team members).

These structures manifests in three behavioral dimensions: *specialization* (recognition of other's expertise), *credibility* (trust in others' expertise), and *coordination* (effective task division) (Lewis & Herndon, 2011). These have been shown to enhance team performance, creativity, and adaptation across various collaborative contexts. For instance, Kanawattanachai & Yoo (2007) found task-oriented communication fosters specialization and credibility, which facilitates coordination. Over time, coordination becomes a key factor in enhancing team performance, while communication frequency loses direct impact. Aissa et al. (2022) found that TMS behaviors associated with specialization and coordination directly impact team creativity. Regarding the role of information technology, Sparrow et al. (2011) found individuals use search engines as external memory aids in a TMS manner, remembering where information can be retrieved rather than the information itself. Choi et al. (2010) found knowledge management solutions positively impact TMS development by supporting meta-knowledge.

He et al. (2022) investigated how TMS address dispersions in functional expertise on individual creativity in information technology project teams. Their study revealed that TMS and geographical dispersion interactively moderate individual-level idea generation and implementation processes in multidisciplinary teams, but in qualitatively different ways. Bienefeld et al. (2023) found that access to AI knowledge leads to increased sharing of suggestions and information between team members. McWilliams & Randolph (2024) found that TMS positively affects trust in AI systems, leading to increased knowledge sharing. Hopf et al. (2024) further highlight that AI systems fulfil sufficient requirements to pose as an entity within a TMS, lifting the previous assumption that only human collaboration can be examined in the TMS context.

In summary, research has identified behavioral dimensions and underlying structures regarding TMS, what fosters their development, and effects on human collaborative work. Studies have demonstrated how technology supports TMS and even established that AI can be incorporated into the TMS discourse. However, literature has not yet shown how TMS processes and structures manifest in practice, nor whether differences between human-only and human-GenAI collaboration warrant necessary changes to theory surrounding TMS in the context of human-AI collaboration.

## 3 Research Design

To examine how TMS manifest in human-GenAI collaboration, and how we can provide a nuanced view on TMS to better represent these socio-technical relationships, we employed a qualitative interpretive approach.

In order to gain lived rich insights (Walsham, 2006) into collaborative experiences between humans and GenAI, we followed Klein & Myers' (1999) principles for interpretive field research and choose semi-structured interviews for data collection.

To ensure that our semi-structured interviews would provide sufficient data to examine TMS manifestations grounded in practice, we followed Eisenhardt's (1989) guidelines and purposefully sampled knowledge workers who regularly engaged with GenAI systems for various knowledge tasks, gaining access to the socio-technical relationships essential for examining the TMS phenomena we set out to investigate.

This informed analysis approach ensured a rigorous qualitative research methodology to then revisit established theory (Gioia et al., 2013) and subsequently provide a more nuanced perspective on TMS in human-GenAI collaboration

The interview protocol was theoretically grounded in TMS literature, with questions designed to uncover human-GenAI dynamics across the three key dimensions of TMS: specialization, credibility, and coordination (Lewis & Herndon, 2011) as well as the underlying memory structures (Hopf et al., 2024). Participants were asked to describe their collaborative experiences with GenAI, focusing on how they recognized and leveraged each other's expertise, experienced trust, and coordinated their collaborative efforts across various knowledge tasks.

For data analysis, we followed Braun & Clarke's (2006) thematic analysis framework, which provides a structured yet flexible approach for identifying patterns within qualitative data. To further ensure methodological rigor, we also employed Alvesson & Sköldberg's (2017) reflexive approach, acknowledging our theoretical lens while remaining open to emergent themes. This analytical strategy included familiarization with interview transcripts, inductive coding of relevant segments, and progressive refinement of codes into broader themes. We then further reviewed these themes through the theoretical lens of TMS, examining where human-GenAI collaboration aligned with or departed from traditional TMS concepts.

This data collection continued until theoretical saturation was reached (Charmaz, 2014), where additional interviews no longer yielded substantive new insights about TMS dynamics in the collaborative arrangements. Theoretical Saturation was reached after conducting and analyzing interviews with the 14 participant listed in Table 1.

**Table 1.** Participants

| ID | Gender | Collaborative experiences with GenAI |
|---|---|---|
| P01 | Male | Text processing, programming, knowledge acquisition |
| P02 | Male | Text processing, programming, knowledge acquisition |
| P03 | Male | Text processing, data analysis |
| P04 | Male | Programming, knowledge acquisition |
| P05 | Female | Text processing, programming, knowledge acquisition |
| P06 | Male | Text processing, knowledge acquisition |
| P07 | Male | Text processing, programming, knowledge acquisition |
| P08 | Male | Text processing, knowledge acquisition, personal assistance |
| P09 | Female | Text processing, knowledge acquisition, personal assistance |
| P10 | Male | Text processing, knowledge acquisitions, problem solving |
| P11 | Male | Data analysis, knowledge acquisition |
| P12 | Female | Text processing, knowledge acquisition, personal assistance |
| P13 | Male | Text processing, programming, knowledge acquisition |
| P14 | Male | Text processing, knowledge acquisition |

## 4 Results

Our interviews conducted with 14 participants lasted between 19 and 89 minutes, with an average of 29 minutes. Participants revealed that integrating GenAI in knowledge tasks has become a daily occurrence, where GenAI is often approached first instead of humans. Eleven participants stated that they engaged with GenAI daily or almost daily, while three reported engagements multiple times a week.

In line with TMS, participants possessed an overall awareness when GenAI is suitable to be integrated in a wide range of knowledge tasks, including text editing, programming, knowledge gathering, and personal assistance. Among the most common tasks were various forms of text processing: "*So I always write down bullet points of the information […] then, after this process is complete, I first sort the points again […] and then I copy and paste that directly into ChatGPT […] then I copy and paste the [results] into the reports, and then I check again whether I like the phrasing, because sometimes they are a bit clunky and mistakes still happen. I then correct those*" (P09). Participants also integrated GenAI into their programming tasks: "*After I had clarified all my questions, I implemented it in an Excel sheet, for example, and then I got an error, then I also ask for feedback and correction of my code, saw the answer, follow-up questions, then understood the code, tried to replicate and implement the code myself, got an error and then questioned further to find the solution to the problem*" (P07). GenAI has also largely replaced traditional search engines for information gathering: "*I've shifted to asking AIs more, because it's simply closer to communicating with a person. You can throw follow-up questions at it when you want more specific answers*" (P04).

Executing the thematic analysis as described in the research design resulted in the identification of the following four majors themes.

### 4.1 Recognized Expertise, but Questioned Credibility

Contrary to the TMS structure where credibility manifests through expertise recognition, participants revealed an ambivalent relationship with GenAI regarding trust. Participants simultaneously attributed expertise to GenAI for specific knowledge tasks while also doubting the results for tasks they had entrusted GenAI with.

Outputs based on content that has already been vetted by participants and passed to GenAI for further processing was generally seen as credible: "*I find improvements of texts that are already clear in terms of content or are asking for improvement in language, I find that's always really great*" (P08). This stands in contrast to tasks that incorporated unknown information provided by GenAI, where participants stated that outputs can't be trusted: "*Especially when it comes to scientific facts, [AI is] always wrong somewhere, and quotes anyway, so I […] critically questioned it*" (P08); "*And currently we have it such that I don't trust the output, well I don't trust it 100 percent in any case*" (P01); "*the AI is very reliable at spreading fake news. Yes, it always acts as if it knows everything, even when it doesn't*" (P02).

As a result participants saw themselves forced to incorporate additional manual verification processes: "*AI also tends to sometimes make things up a bit, and this experience has shown that for me in the professional world, the direct one-to-one application of AI is simply not feasible, and there is still a certain amount of effort required for verificatio*n" (P10). When working with human collaborators, participants didn't feel the need to verify information due to an inherent trust, especially toward those in higher positions that were associated with greater expertise: "*humans are reliable, [I] would describe the qualitative difference or the confidence in the output as a bit better*" (P07); "*and with my supervisor I'm much more like, oh yes, okay, if he says it like that, then it must be that way, because he is much, much smarter than I am*" (P05).

### 4.2 Human Coordination facing GenAI's lack of Initiative

Whereas TMS describes a distributed system where members actively coordinate based on recognized expertise domains, participants revealed an asymmetric process for distributing labor with GenAI. While cognitive tasks were divided between human and GenAI, this division occurred through hierarchical assignment rather than drawing on each other's strengths through mutual exchange of ideas and expertise.

As one participant stated, "*with AI, I am the initiator*" (P01) while another reinforced this dynamic by explaining "*with AI it's always like this, the AI is very passive [...], actually only ever reacts to what I give it*" (P04). This relationship was frequently framed in explicitly hierarchical terms through workplace metaphors. Participants used descriptions like "*boss-employee [dynamic]*" (P01), "*assistant*" (P08), and "*service provider*" (P05), characterizing GenAI as a subordinate team member.

The hierarchical dynamic created a workflow where participants issued directives and AI executed without contributing desired direction: "*when it comes to AI, I still feel like [the AI] might know a bit more, but I'm still the one who decides […], because in the end it's up to me what I accept from [the AI] or what not*" (P05). This contrasted with human collaboration, which was described as more balanced: "*A human is proactive, has more initiative and also approaches me*" (P04).

As a result, GenAI rarely challenged thinking or pushed conversations in new directions: "*I just get the output, but then there isn't a follow-up question or a different opinion coming back where you can discuss things further*" (P01). With humans, participants could "*hack out a plan*" (P07) together: "*You discuss and get from one topic to another, and inspire each other*" (P12).

### 4.3 Asymmetric and Volatile Expertise Recognition

Unlike traditional TMS, where collaborators maintain an awareness of each other's knowledge domains for efficient coordination, participants revealed that GenAI operated with no meaningful understanding of human collaborators' expertise or skills.

Participants reported they held awareness of the GenAI's capabilities, while GenAI lacked the awareness of participants' background and expertise: "*the AI usually doesn't have that much contextual knowledge about me, but I have more contextual knowledge about the AI. And with humans, it's more reciprocal*" (P03); "*I don't have the feeling* *that the system knows me or my behavior*" (P10). This imbalance manifested in GenAI's inability to calibrate responses to match participants expertise levels: "*I don't believe that AI says, yes okay, on this topic I don't need to give him as much context, because he already knows a lot about it*" (P01);

However, participants explained that AI was able to develop a temporary recognition of participants expertise within a temporary context. This understanding emerged through ongoing interaction: "*And when [AI] has been well fed with information in a chat, I have the feeling that the AI definitely has more contextual knowledge*" (P03). However, this recognition remained strictly confined to individual sessions. GenAI's ability to understand a participant's background, skills, and knowledge disappeared when starting a new interaction: "*As soon as I go to a different chat, it's lost again. Then I have to start from the beginning again. So it's always like a restart*" (P01).

Participants identified this limitation as a fundamental disadvantage compared to human collaboration, where recognition of expertise persists across interactions: "*And that's an ability that humans have. When I talk with colleagues or people who already know me, they know my attitudes toward things, they know what I work on, they basically know my knowledge base. And with them, you can start at a completely different level right away*" (P01). According to participants, this persistent memory enables humans to act proactively: "*AI is, as mentioned, very limited, simply in the sense that it's only in this specific context [...]. But with humans, it's more like they have their own agenda [...] if I'm the person with relevant expertise, they come to me on their own*."

### 4.4 Unilateral Resource Integration

While TMS emphasize external storage as an extension of group memory where documents function as shared cognitive artifacts accessible to all collaborators, participants described an asymmetric approach do document-mediated collaboration with GenAI, which differed from their traditional human-human collaborative practice.

With GenAI as a collaborative partner, sharing often only took place in one direction, rather than through truly collaborative documents which both parties use to actively store, share and process information. Knowledge tasks often included integrating external information through uploading files: "*upload those to [the AI] for example and tell it, let's summarize this briefly*" (P12); "*they are sent directly as PDF, bam, sent over*" (P02). Manual pasting of text was another strategy: "*so I always work in a way that I just send over code snippets, [...] and [the AI] then gives me a revised version back [...] my workflow [...] Copy, paste, send it*" (P02). One participant with a background in psychology described sharing clinical notes: "*so while I conduct these diagnostic conversations, I'm taking notes the whole time [...]. So I always write down bullet points, that is, the information that I have, so to speak, gathered from the conversation [...] and then I copy and paste that directly into [the AI]*" (P09).

This one-sided flow of information stands in contrast to human-human collaboration, where participants recognized the value of mutual engagement with shared documents. As one participant noted: "*If I load a document now, it's a bit one-sided in the sense that the AI explains something to me, but I can't really discuss it properly with it. But with humans, in my opinion, the ability to discuss the material again is better*" (P04).

# 5 Discussion

## 5.1 Manifestation of TMS in Human-GenAI Collaboration

Addressing our research question on how TMS manifest in human-GenAI collaborations, our analysis reveals both parallels and deviations across the three key TMS dimensions of *specialization*, *credibility* and *coordination*.

Concerning *specialization*, the recognition of expertise with GenAI remained asymmetric. Humans developed awareness of GenAI specializations in line with TMS, but GenAI failed to recognize human expertise domains, failing Fügener et al.'s (2022) collaboration conditions. There were exceptions where GenAI demonstrated a limited grasp of human expertise, but this recognition was confined to individual collaborative sessions rather than persisting across interactions. This limitation forced knowledge workers to repeatedly establish context in each new interaction, creating inefficiencies compared to human collaborators who can immediately engage at appropriate expertise levels due to their persistent memory of professional backgrounds.

Regarding *credibility*, participants displayed an ambivalent trust relationship towards GenAI, simultaneously attributing expertise while also doubting its results, echoing the ‘algorithm aversion’ phenomenon (Dietvorst et al., 2018). This differs from TMS, where collaboration stems from pre-existing trust. It’s noteworthy that interaction with GenAI was sought amid those trust concerns. This ambivalence led participants to develop additional verification processes for GenAI outputs, creating a careful form of collaboration that required continuous assessment rather than the established confidence typically found within TMS structures.

For *coordination*, human-GenAI collaboration was characterized by hierarchical rather than mutual dynamics, with humans directing and GenAI acting without meaningful initiative, further exemplifying the ‘irony of automation’ (Rowe et al., 2024). Interaction remained predominantly unidirectional, with AI functioning as a passive responder rather than an active knowledge-seeking participant. Additionally, unlike human collaborators which jointly access and contribute to shared knowledge stores, GenAI engaged minimally with external information sources, creating a one-way knowledge flow. This passive coordination pattern limited collaborative innovation, as the spontaneous exchange of ideas and creative evolution of concepts that characterizes balanced collaboration remained largely absent and was depended on human initiation.

## 5.2 A Nuanced Perspective on TMS in Human-GenAI Collaboration

Addressing the second part of our research question on how a nuanced view on TMS theory can cover for these collaborative relationship changes, our analysis suggests theoretical adaptations to memory structures within TMS that reflect the unique dynamics of human-GenAI knowledge collaborations, as depicted in Figure 1.

Expertise recognition in TMS functions as a mental directory for identifying knowledgeable collaborators. While TMS theory would predict rejection of unreliable sources, participants actually maintained GenAI as a knowledge partner despite trust concerns, simply adding verification steps. This suggests an evolution in TMS meta-memory structure, where expertise recognition includes trust markers that trigger verification behaviors during collaboration, creating a spectrum-based rather than categorical approach to collaborative information processing with variably reliable partners.

Further, according to TMS, while humans maintain robust mental directories of collaborators' expertise, GenAI demonstrated fundamental limitations in this architecture. Based on GenAI's passive approach to collaboration, where it fails to initiate collaborative behaviors and utilize shared storage only when explicitly human-directed, we diagnose the absence of a persistent metamemory structure for GenAI compared to human counterparts within TMS. GenAI simultaneously maintains extensive knowledge but lacks the metamemory required to store domain-expertise associations about potential collaborators, preventing it from seeking information or initiating other forms of collaborative behaviors. We therefore suggest the existence of another memory type within the personal memory of GenAI: conversation-bound temporary memory. While GenAI lacks persistent meta-memory for expertise recognition, it can temporarily store both knowledge and metamemory components within its current conversation context. This temporary memory allows GenAI to reference both task-related knowledge and expertise recognition information shared earlier in the same session, creating a limited simulation of meta-memory function within the conversation boundary. However, these temporary memories remain isolated and volatile, disconnected from GenAI's core knowledge base and unable to persist across separate interactions. This further distinguishes GenAI's memory architecture from the human counterpart.

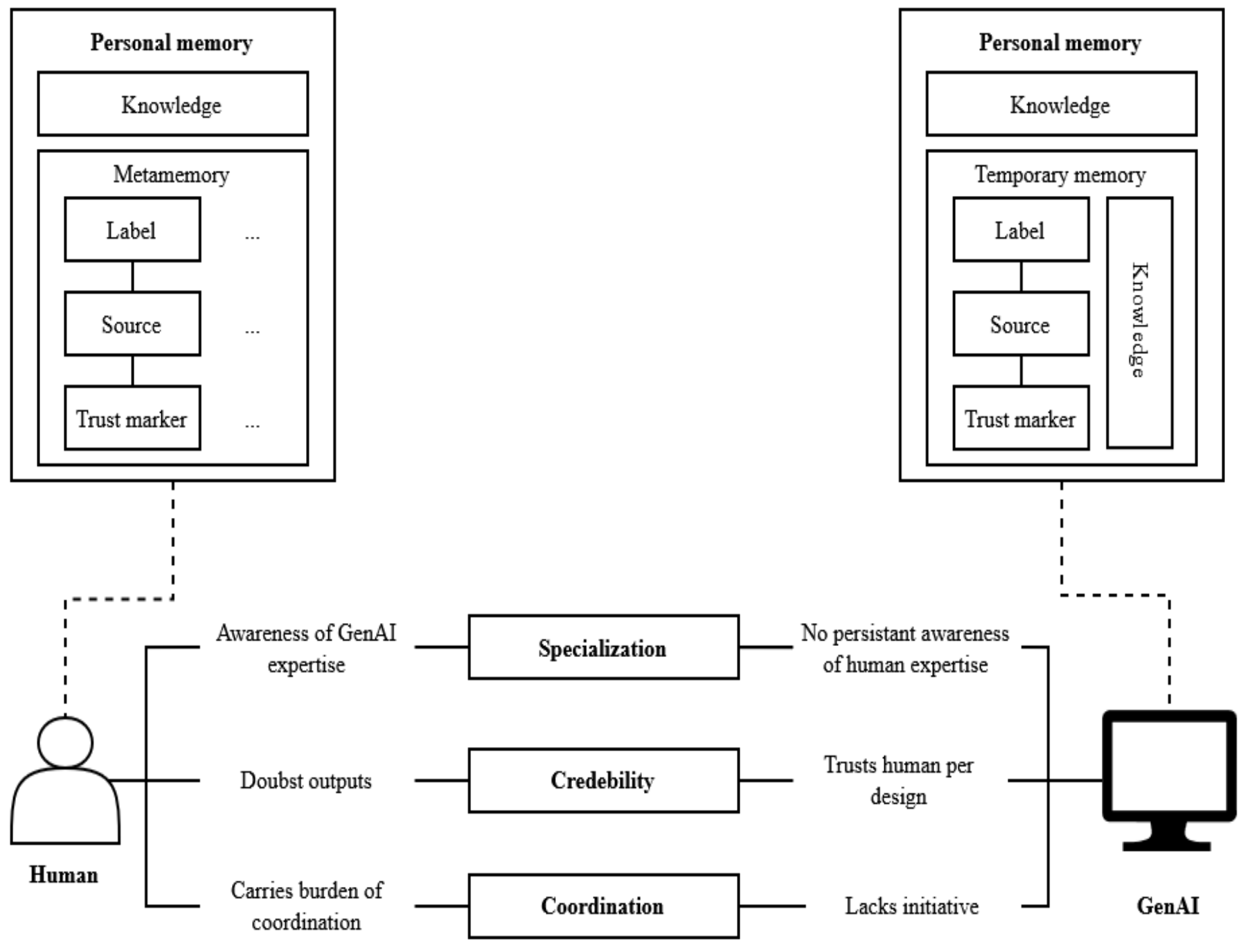

**Figure 1.** TMS manifestation and adapted memory structure in human-GenAI collaboration

### 5.3 Implication for theory and practice

Theory must transform TMS from a purely sociological construct into a comprehensive socio-technical framework by systematically revisiting each memory type and transactive process, while being open to discover completely new ones. This theoretical evolution requires mapping traditional socio-cognitive TMS components to appropriate socio-technical equivalents that account for GenAI's distinct technical architecture compared to the social nature of humans. Such reconceptualization will not only advance TMS theory but also provide crucial guidance for designing GenAI systems that better align with human collaborative cognition patterns. Our paper contributes to this discourse by starting to analyze the underlying TMS structures in human-GenAI contexts and provide first directions regarding the required mapping process.

Our findings also urge practitioners to reassess TMS to adapt structures for effective human-GenAI collaboration. Current collaborations operate sub-optimally due to socio-technical dynamics that traditional TMS fail to capture. Information systems professionals should design suitable mechanisms for persistent expertise recognition, offer appropriate verification mechanisms, and develop features countering GenAI's passive nature, by providing a metamemory structure for GenAI. Interfaces supporting bidirectional knowledge exchange and coordination mechanisms that complement GenAI's strengths could significantly improve collaborative outcomes, bridging theory and practice in human-GenAI knowledge partnerships.

## 6 Conclusion

This research investigated TMS in human-GenAI collaboration through qualitative interviews with knowledge workers, revealing disruptions to traditional TMS theory due to trust issues, hierarchical control maintenance, limitations in GenAI's proactive behavior, and absent awareness of human expertise, all impeding optimal collaboration. Our paper thus proposes adaptations to traditional TMS theory: augmentation of human meta-memory with trust markers, acknowledgment of GenAI's absent metamemory, and identification of GenAI's temporary memory, aiding future GenAI system design.

The study's limitations include its sample size of 14 participants, which may not capture the full spectrum of human-GenAI collaborative experiences. Additionally, to adequately examine TMS structures, we selected participants familiar with GenAI who engage it regularly for knowledge tasks to gain initial insights, though it's possible that novices could reveal additional facets in first contact scenarios. The self-reported nature of the data also relies on participants' subjective perceptions, which may not fully capture the objective reality of these collaborative interactions.

Future research should further examine the memory structures and transactive process to aid the TMS transition from a purely socio-cognitive to a socio-technical perspective. Studies should start to implement mechanisms for our identified memory adaptions in TMS and examine the impacts, before then investigating human-GenAI TMS in longitude studies with GenAI veterans as well as novices.